\documentclass[twocolumn,prb,superscriptaddress,noshowpacs,epsf]{revtex4}
\usepackage{graphicx}

\begin{document}
\title{Correlation-induced suppression of decoherence in 
capacitively coupled Cooper-pair boxes}
\date{\today}
\author{J. Q. You}
\affiliation{Frontier Research System, The Institute of Physical and Chemical
Research (RIKEN), Wako-shi 351-0198, Japan}
\affiliation{Department of Physics and Surface Physics Laboratory (National
Key Laboratory), 
Fudan University, Shanghai 200433, China}
\author{Xuedong Hu}
\affiliation{Frontier Research System, The Institute of Physical and Chemical
Research (RIKEN), Wako-shi 351-0198, Japan}
\affiliation{Department of Physics, University at Buffalo, SUNY, 
Buffalo, NY 14260-1500, USA}
\author{Franco Nori}
\affiliation{Frontier Research System, The Institute of Physical and Chemical
Research (RIKEN), Wako-shi 351-0198, Japan}
\affiliation{Center for Theoretical Physics, Physics Department, Center for
the Study of Complex Systems, The University of Michigan, Ann Arbor, MI
48109-1120, USA}

\begin{abstract}
Charge fluctuations from gate bias and background traps severely limit the
performance of a charge qubit in a Cooper-pair box (CPB).  Here we present an
experimentally realizable method to control the decoherence effects of these
charge fluctuations using two strongly capacitively coupled CPBs.  This
coupled-box system has a low-decoherence subspace of two states.  
Our results show that the inter-box Coulomb correlation can help significantly
suppress decoherence of this two-level system, making it a promising
candidate as a logical qubit, encoded using two CPBs.
\end{abstract}
\maketitle

\section{Introduction}

Various superconducting nanocircuits have been proposed as quantum bits
(qubits) for a quantum computer.~\cite{NEC1,MIT1,SUNY,HAN,MSS1}  In the meantime,
it has long been recognized that background charge fluctuations can severely
limit the performance of microelectronic devices, particularly those based on
the manipulation of electrical charge, such as single electron
transistors~\cite{SET} and superconducting Cooper-pair boxes
(CPBs).~\cite{NEC2,PALA,SAC2}  The struggle to suppress or even eliminate
noise from charge fluctuations in superconducting devices has been a
prolonged battle with limited success.  Here, instead of focusing on
perfecting materials, we propose an alternative experimentally-realizable
method to suppress the effects of these charge fluctuations using two
strongly (capacitively) coupled CPBs.

Cooper-pair boxes are one of the prominent candidates for qubits in a
quantum computer.  Recent experiments~\cite{NEC3} have revealed quantum
coherent oscillations in two CPBs coupled capacitively and demonstrated the
feasibility of a conditional gate as well as creating macroscopic entangled
states.  
Scalable quantum-computing schemes (see, e.g., Ref.~\onlinecite{YTN}) have also been 
proposed based on
charge qubits.  Clearly, effective suppression of charge noise is essential
to the practical implementation of scalable quantum computing in a
charge-based scheme.  It has been shown~\cite{SAC2} that while operating at
the degeneracy point (where the two lowest charge states have the same energy
in the absence of Josephson coupling), the charge qubit
has a long decoherence time of $\tau\approx 500$ ns.  However, when the charge
qubit is operated away from the degeneracy point, it experiences strong
dephasing by the charge fluctuations, and the decoherence time of the system
is greatly reduced.~\cite{NEC2,SAC2,COTT}

Two separate CPBs generally experience uncorrelated charge fluctuations as
they are most strongly affected by their own gate biases and the nearest
fluctuating charge traps.  However, if the two boxes are strongly coupled
capacitively (with no tunnel coupling so that the two-box states do not
approach those of a single large box in the strong coupling limit), the
fluctuations affecting one box will affect the other through Coulomb
interaction.  
In the limit of extremely strong inter-box coupling (corresponding to a very
large mutual capacitance between the two CPBs), the two boxes would
experience an identical charge environment, so that, in principle, a
decoherence-free subspace~\cite{ZR,DUAN} could be established for coupled-box
states.  However, 
in reality this limit involves many degenerate charge states for the
electrostatic energy of the coupled boxes, so that logical qubit encoding is
impossible. 
Can we still achieve a decoherence-suppressed logical qubit in two
capacitively coupled boxes?  Below we show that there exists an intermediate
parameter regime where a strong inter-box Coulomb correlation induces a
significant suppression of decoherence in certain two-box states, so that
considerable benefit can be reaped by encoding a logical qubit in terms of
these states.

\begin{figure}
\includegraphics[width=3.2in,bbllx=65,bblly=250,bburx=515,bbury=766]
{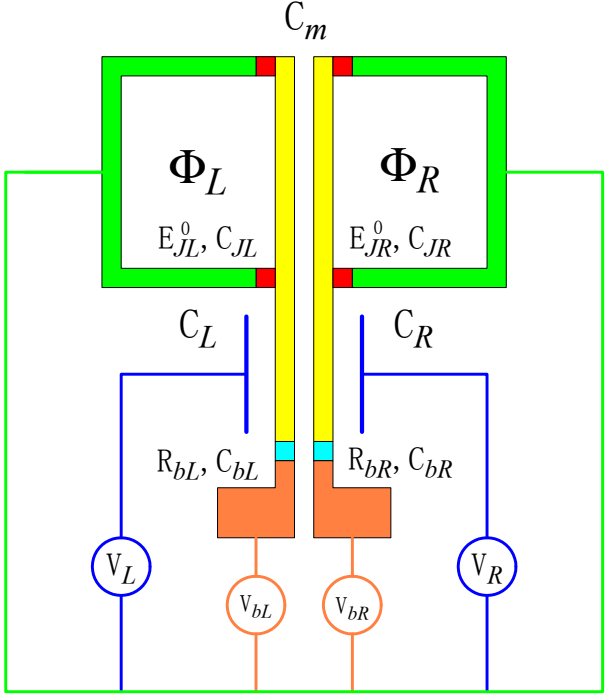}
\caption{(Color online) Strongly coupled Cooper-pair boxes.  A bias voltage
$V_i$ is applied to the $i$th charge box through a gate capacitance $C_i$,
and a symmetric dc-SQUID (with Josephson coupling energy $E^0_{Ji}$ and
capacitance $C_{Ji}$ for each junction) is coupled to the box.  Also, each
box is connected to a detector via a probe junction (or a less invasive point
contact).  When a measurement is performed, the probe junction is biased with
an appropriate voltage $V_{bi}$. The two boxes are closely-spaced long
superconducting islands with sufficiently large mutual capacitance $C_m$, and
the barrier between them is strong enough to prohibit the inter-box
Cooper-pair tunneling.}
\end{figure}

\section{Characterization of two coupled Cooper-pair boxes}

Consider two capacitively coupled CPBs (see Fig.~1).  Each CPB is individually 
biased by an applied gate voltage $V_i$ and coupled to the leads by a symmetric
dc-SQUID. The dc-SQUID is pierced by a magnetic flux $\Phi_i$, 
which provides a tunable effective Josephson coupling 
\begin{equation}
E_{Ji}(\Phi_i) = 2E^0_{Ji} \cos\left({\pi\Phi_i\over\Phi_0}\right), 
\end{equation}
where $\Phi_0=h/2e$ is the flux quantum.  
The system Hamiltonian is
\begin{eqnarray}
H_S & = & \sum_i [E_{ci}(n_i-n_{xi})^2 - E_{Ji}(\Phi_i) \cos\varphi _i]
\nonumber \\
& & + E_m(n_L-n_{xL})(n_R-n_{xR}) \,,
\end{eqnarray} 
with $i=L,R$ for left and right.  Here the charging energy $E_{ci}$ of the
$i$th superconducting island and the mutual capacitive coupling $E_m$ are
given by~\cite{MB} 
\begin{eqnarray}
&& E_{ci}={2e^2C_{\Sigma j}\over\Lambda}\,,\nonumber \\
&& E_m = {4e^2C_m\over\Lambda}\,, 
\end{eqnarray}
with $\Lambda$ given by 
\begin{equation}
\Lambda=C_{\Sigma i}C_{\Sigma j}-C_m^2\,, 
\end{equation}
where 
\begin{equation}
C_{\Sigma i}= C_m + C_i + C_{Ji} 
\end{equation}
is the total capacitance of the $i$th island.  The
offset charge is 
\begin{equation}
2en_{xi} = Q_{Vi}+ Q_{0i}\,, 
\end{equation}
where $Q_{0i}$ is the
background charge, and 
\begin{equation}
Q_{Vi}=C_i V_i + C_{bi} V_{bi}
\end{equation}
is induced by both
the gate voltage $V_i$ and the probe voltage $V_{bi}$.  The average phase
drop $\varphi_i$ across the two Josephson junctions in the dc-SQUID is
conjugate to the Cooper pair number $n_i$ on the box.  Both CPBs operate in
the charging regime $E_{ci}\gg E_{Ji}$ and at low temperatures $k_B T \ll
E_{ci}$.  The states of the two coupled boxes can thus be expanded on the
basis of the charge eigenstates $|n_L n_R\rangle \equiv |n_L \rangle
|n_R\rangle$.

When the two CPBs are strongly coupled, the total Hamiltonian can be rewritten
in terms of the total charge on the coupled boxes and the charge difference
across the boxes. Assuming, for simplicity,
\begin{equation}
C_{\Sigma L}= C_{\Sigma R} = C_{\Sigma}\,,
\end{equation}
so that 
\begin{eqnarray}
&& E_{cL}=E_{cR}=E_c \, ,\nonumber\\ 
&& E_{JL}=E_{JR}=E_J \, , 
\end{eqnarray}
we have
\begin{eqnarray}
H_S & = & \left(E_c - \frac{\Delta E}{2}\right) \left(n_L + n_R 
- n_{xL} - n_{xR}\right)^2 \nonumber \\
& & + \frac{\Delta E}{2} \left(n_L-n_R-n_{xL}+n_{xR}\right)^2 \nonumber\\ 
& & - E_J( \cos\varphi _L+\cos\varphi _R)\,,
\end{eqnarray}
where 
\begin{equation}
\Delta E \;=\; {2e^2\over C_m+C_{\Sigma}} 
\;=\; E_c - {1\over 2}E_m \;>\;0 \,.
\end{equation}
Notice that when $C_m\gg C_i$, and $\ C_{Ji}$, 
$E_c$ essentially represents the charging energy of
individual Josephson junctions, while $\Delta E$ represents the charging 
energy of the large capacitor $C_m$, so that $\Delta E \ll E_c$.

\begin{figure}
\includegraphics[width=3.4in,bbllx=83,bblly=277,bburx=501,bbury=647]
{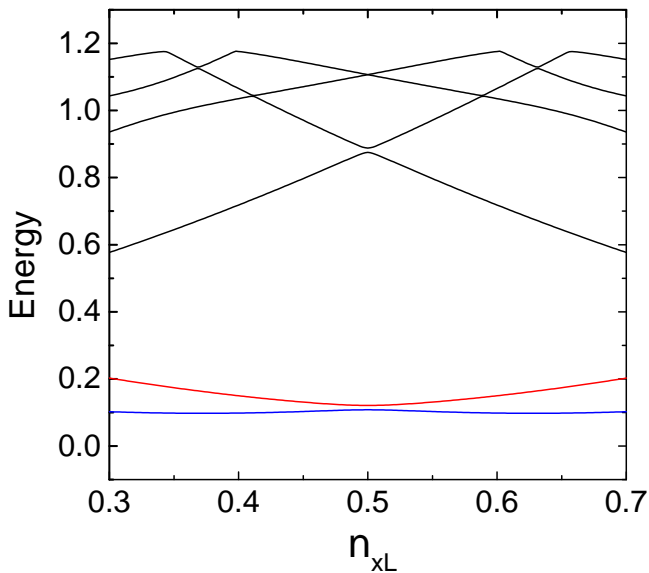}
\caption{(Color online) Dependence of the energy levels of the coupled-box
system on the reduced offset charge $n_{xL}$ for $n_{xR}=0.5$.  Here $\Delta
E_i=E_{ci}/4$, and $E_{Ji}=E_{ci}/10$, with $i=L,R$.  The energy is in units
of $E_{c}$.  We choose 10 two-box basis states $|m,n\rangle$ that have the
lowest electrostatic energy.  The two lowest levels remain nearly unchanged
in the vicinity of the degeneracy point $(n_{xL},n_{xR})=({1\over 2},{1\over
2})$.
}
\end{figure}

At the double degeneracy point of
$(n_{xL},n_{xR})=({1\over 2},{1\over 2})$, 
the two lowest energy states
are given by 
\begin{equation}
|\pm\rangle = {1\over \sqrt{2}}(|01\rangle \pm |10\rangle) + 
|\delta \psi_{\pm}\rangle \, , 
\end{equation}
where 
\begin{equation}
|\delta \psi_{\pm} \rangle = O\left({E_J\over E_m}\right)
[\,\alpha_{\pm} (|00\rangle \pm |11\rangle) + \cdots\,] \, , 
\end{equation}
with a splitting of $E_J^2/2(E_c-\Delta E)$ (see Fig.~2).  
The symmetry in these states indicates that they are well insulated from pure
dephasing and relaxation due to charge noise, as we will show below.  It is
thus quite natural to adopt these two coupled-box states $|\pm\rangle$ to
encode a logical qubit.  Below we calculate the dephasing and relaxation
properties of the $|\pm\rangle$ states and discuss how they can be coherently
manipulated.

\section{Correlation-induced coherence-preserving subspace} 

To clarify
the origin of the correlated environments for the two coupled CPBs, we study
the fluctuations~\cite{NOTE} of the reduced offset charge $n_{xi}$, which
could originate from the gate voltage $V_i$, probe voltage $V_{bi}$, and
background charge $Q_{0i}$.  The interaction Hamiltonian between the
charge noise and the coupled CPBs takes the form
\begin{eqnarray}
H_I & = & -2 \left(E_c - \frac{\Delta E}{2}\right) \left(n_L + n_R\right)
\left( \delta n_{xL} + \delta n_{xR}\right) \nonumber \\
& & - \Delta E (n_L - n_R) (\delta n_{xL} - \delta n_{xR}) \,.
\label{eq:H_I}
\end{eqnarray}

We can use the language of a two-level system to describe each of the CPBs 
around the degeneracy point 
$$(n_{xL},n_{xR})=\left({1\over 2},{1\over 2}\right),$$ 
and rewrite the system Hamiltonian in terms of the Pauli matrices:  
\begin{eqnarray}
H_S \!&\! =\! &\! \sum_i H_i + {1\over 4}E_m \sigma_{zL}\sigma_{zR}\,,\\
H_i \!&\! =\! &\! [\varepsilon_i(n_{xi}) + \varepsilon_m(n_{xj})] \sigma_{zi}
- {1\over 2} E_{Ji}(\Phi_i) \sigma_{xi}\,, \nonumber
\end{eqnarray}
with
$|\!\uparrow\rangle_i \equiv |0\rangle_i$,
$|\!\downarrow\rangle_i \equiv |1\rangle_i$, and $i,j=L,R$ ($i\ne j$).
Here 
\begin{eqnarray}
&&\varepsilon_i(n_{xi})=E_{ci}(n_{xi}-{1\over 2}),\nonumber\\
&&\varepsilon_m(n_{xj})={1\over 2}E_m(n_{xj}-{1\over 2}),
\end{eqnarray} 
The interaction Hamiltonian Eq.~(\ref{eq:H_I}) between the CPB system and the
environment can now be projected onto the single-box two-level basis:
\begin{eqnarray}
H_I & = & \sum_{i} (E_{ci} \delta n_{xi} + {1\over 2} E_m\delta
n_{xj}) \sigma_{zi}\,, \nonumber \\
& = & E_c (\sigma_{zL} + \sigma_{zR})(\delta n_{xL} + \delta n_{xR}) 
\nonumber \\
& & -\Delta E (\sigma_{zL} \delta n_{xR} + \sigma_{zR} \delta n_{xL}) \,.
\label{eq:H_I_delta}
\end{eqnarray}
with $i,j=L,R$, and $i\ne j$.  Though each CPB is directly coupled to its own
charge environment, the inter-island Coulomb interaction in terms of $E_m$
ensures that the environment is partly shared between the two islands,
causing the CPBs to experience correlated noises.  Indeed, notice that in
Eq.~(\ref{eq:H_I_delta}) the first part of the Hamiltonian should not lead to
dephasing between the $|\pm\rangle$ states since it affects both identically. 
When each of the two environments is modeled by a thermal bath of simple
harmonic oscillators described by the annihilation (creation) operator
$b_{ji}$ ($b_{ji}^{\dag}$), the Hamiltonian of the whole system, including
the two baths, is
\begin{eqnarray}
&&H=H_S+H_B+H_I\,,\nonumber\\
&&H_B=\sum_n(\hbar\omega_{nL}b_{nL}^{\dag}b_{nL}
+\hbar\omega_{nR}b_{nR}^{\dag}b_{nR})\,,\\
&&H_I = {1\over 2}\sum_{i} 
\left(\sigma_{zi}+{E_m\over 2E_{ci}}\sigma_{zj}\right)X^{(i)},\nonumber
\end{eqnarray}
with $i,j=L,R$ ($i\ne j$).  Here, the bath operator 
\begin{equation}
X^{(i)}=2E_{ci}\delta n_{xi}
\end{equation} 
is given by 
\begin{eqnarray}
X^{(i)}\!&\!=\!&\!\sum_n\lambda_{ni}x_n^{(i)}\nonumber\\
&\!\equiv\!&\! \sum_n \hbar K_n^{(i)}(b_{ni}^{\dag}+b_{ni})\,, 
\end{eqnarray}
with 
\begin{equation}
K_n^{(i)} = {\lambda_{ni} \over \sqrt{2m_{ni} \hbar \omega_{ni}}}\,,
\end{equation}
where $m_{ni}$ and $\omega_{ni}$ denote the mass and frequency of the $n$th harmonic 
oscillator in the bath coupled to the box $i$, while $\lambda_{ni}$ characterizes 
the coupling strength between the $i$th oscillator and the box.

We first focus on pure dephasing between CPB states with $E_{Ji}(\Phi_i)=0$,
which can be solved analytically.~\cite{DUAN} 
For correlated noises studied here, the reduced off-diagonal density matrix
elements for the two lowest energy eigenstates of the coupled-box system
decay as
\begin{equation} 
\rho_{ab} \;\sim\; \exp[-\eta(t)] \, , 
\end{equation}
where the damping factor is given by
\begin{equation}
\eta(t) = \sum_{i=L,R} \left( 1 - {E_m\over 2E_{ci}} \right)^2 
\Gamma_i(t) \,,
\label{eq:tau_Em}
\end{equation}
and $a,b=+,-$ ($a\ne b$). Here $+$ and $-$ denote the two lowest eigenstates.
Also,
\begin{equation}
\Gamma_i(t)={1\over \pi}\int_0^{\infty} d \omega S_i(\omega)
\left({{\sin(\omega t/2)}\over \omega/2}\right)^2,
\end{equation}
and the power spectrum of the $i$th bath is 
\begin{equation}
S_i(\omega) = J_i(\omega) \coth \left({\hbar \omega \over 2k_BT}\right),
\end{equation} 
where 
\begin{equation}
J_i(\omega) = \pi \sum_n[K_n^{(i)}]^2 \delta
(\omega-\omega_{ni})\,.  
\end{equation}
When $t \rightarrow \infty$, $\Gamma_i(t)$ tends
to $t\;S_i(\omega)|_{\omega\rightarrow 0}$.  For the symmetric case we are
considering, 
\begin{equation}
\Delta E = E_c - {1\over 2}E_m\,, 
\end{equation}
so that
\begin{equation}
\eta(t) = \left({\Delta E \over E_c}\right)^2 \sum_{i=L,R}
\Gamma_i(t)\,.
\label{eq:eta(t)}
\end{equation}
In the limit of strong inter-box coupling, $\Delta E \ll E_c$, pure dephasing
can then be strongly suppressed as compared to a single CPB.  For example, for
$\Delta E=E_c/10$, the prefactor takes the value $1/100$, so that the
dephasing time is two orders of magnitude longer than when the boxes are only
weakly coupled.  This is in strong contrast to the corresponding single CPB
expression for pure dephasing 
\begin{equation}
\eta(t)=\Gamma(t)\,.
\end{equation}  
For the $1/f$ noise
arising from the background charge fluctuations, the power spectrum is
\begin{equation}
S_{fi}(\omega)=\left(2E_{ci}\over\hbar e\right)^2{\alpha_i\over\omega}, 
\end{equation}
and $\Gamma_i(t)$
can be written as~\cite{NEC2} 
\begin{equation}
\Gamma_i(t)={1\over \pi}\int_{\omega_c}^{\infty} d
\omega \; S_{fi}(\omega) \left({{\sin(\omega t/2)}\over \omega/2}\right)^2,
\end{equation}
where the cutoff frequency $\omega_c\ll 1/\tau$, with $\tau$ being the
decoherence time of the system.

The above calculation focuses on the pure dephasing of the coupled boxes and
is applicable to parameter regimes away from the double degeneracy point
$$(n_{xL},n_{xR})=\left({1\over 2},{1\over 2}\right)\,.$$  
At the degeneracy point, we can
estimate the effects of the charge fluctuations by directly projecting  
the system-environment coupling Hamiltonian (\ref{eq:H_I}) into the
$|\pm\rangle$ basis.  The matrix elements are
\begin{eqnarray}
\langle+|H_I|+\rangle\!&\! =\!&\! \langle-|H_I|-\rangle \nonumber\\
&\!=\!&\! -2 \left( E_c
-{1\over 2}\Delta E \right) \left(\delta n_{xL} + \delta n_{xR} \right) \,,
\nonumber \\
\langle-|H_I|+\rangle\!&\! =\!&\! \Delta E \left( \delta n_{xL} - \delta
n_{xR}
\right) \,.
\label{eq:H_I_pm}
\end{eqnarray}
From the first equation of Eq.~(\ref{eq:H_I_pm}), 
\begin{equation}
\langle+|H_I|+\rangle - \langle-|H_I|-\rangle =0, 
\end{equation}
so that there is no pure dephasing between
$|+\rangle$ and $|-\rangle$ states as the charge fluctuations affect both
identically.  On the other hand, the second equation of Eq.~(\ref{eq:H_I_pm}),
i.e., the transition matrix element,
dictates that charge fluctuation does lead to relaxation between these two
states.  Using the spin-boson model above, one can calculate this transition
rate straightforwardly.  Here we emphasize that compared to a single CPB, the
system-reservoir interaction strength is $\Delta E$ instead of $E_c$, just
like in Eq.~(\ref{eq:eta(t)}) for pure dephasing.  Furthermore, charge
fluctuations that couple to the two boxes equally will not lead to relaxation
because the coupling here is proportional to $\delta n_{xL} - \delta n_{xR}$.

In short, a pair of capacitively coupled CPBs can have strongly suppressed
pure dephasing and relaxation around the degeneracy point
because of the reduced interaction strength.  
Therefore $|\pm\rangle$ are perfect candidates to encode a logical qubit. 
Moreover, as shown in Fig.~2, the two lowest levels are very separated from 
the higher levels in the coupled CPBs and the leakages from the qubit states 
to the higher-level states can be negligibly small. 
In contrast, for the single CPB qubit in the charge-flux regime where 
the charging energy is reduced,~\cite{SAC2}
the two lowest levels are not well separated from the higher 
levels~\cite{COTT} and appreciable leakages are expected.

\section{Discussion and conclusion}

Coherence-preserving quantum states can be prepared as follows.  First,
consider an initial point on the $n_{xL}$--$n_{xR}$ plane close to $(0,0)$. 
Here the system ground state is $|00\rangle$.  Then, shifting adiabatically
(e.g., along the $n_{xL}=n_{xR}$ direction) to the region around the
degeneracy point, we arrive at the coherence-preserving ground state
$|+\rangle$.  Now, using a two-frequency microwave to interact with the
system for a period of time (basically a Raman process), as in the case of
trapped ions,~\cite{iontrap} one can obtain any superposition of
$|\pm\rangle$ states, so that an arbitrary single qubit operation is
feasible. 
Readout of the logical-qubit states can be achieved by various approaches. 
For instance, one can rotate the logical qubit states to the charge
eigenstates $|01\rangle$ and $|10\rangle$, so that simple charge
detection using, for example, single electron transistors, can
determine the state of the coupled CPBs.  In the case of
probe junction detection (Fig.~1 and Ref.~\onlinecite{NEC3}), when appropriate bias
voltages $V_{bi}$ are applied to the probe junctions, the measured current
$I_i$ through the $i$th probe junction is proportional to the probability for
the $i$th box to have a Cooper pair in it.


Decoherence in two coupled qubits~\cite{GOV,STO} and during a conditional 
gate~\cite{THO} have attracted much attention recently.  It has been shown
that a decoherence-free subspace exists for two physical qubits coupled to
the same bath.~\cite{STO}  Recently, Zhou{\it~et~al}.~\cite{Zhou}
proposed an encoded qubit using a pair of closely spaced CPBs sharing a
common lead, and the two boxes were assumed to couple to an identical bath. 
In their proposed setup, fluctuations originating from the gate voltage may
be identical because of the common lead.  
However, the background charge fluctuations~\cite{NEC4} cannot be so since
these fluctuations originate from local charge traps near each box.  As shown
in Eq.~(\ref{eq:H_I_delta}), an identical bath could only be achieved in the presence
of inter-box interaction and in the limiting case of $E_{ci} = {1\over 2}
E_m$.  Unfortunately, at this limit, the two-level-system description for the
individual CPB breaks down.  Thus the proposed ideal single-bath scenario can
never be achieved in the presence of background charge fluctuations. 
Nevertheless, as shown in our study of the coupled CPBs here,
though the ideal single-bath case cannot be realized to obtain a
decoherence-free subspace, the strong inter-box coupling does enable a
coherence-preserving logical qubit where the correlated baths lead
to suppression of decoherence in the coupled CPBs.   

We emphasize that the encoding idea presented here is based on actively
employing the inter-box interaction to correlate different environments
experienced by the individual physical qubits. 
This goes beyond the decoherence-free subspace concept, where symmetry alone
is used (passively) to combat noise from a naturally existing common
environmental reservoir to all the qubits.

In conclusion, we have shown that in two strongly capacitively coupled CPBs,
the charge fluctuations experienced by the two boxes are strongly correlated.
The inter-box Coulomb correlation creates a two-box subspace of two states in
which pure dephasing and relaxation are strongly suppressed due to the
correlated noises.  These two coupled CPBs can therefore be used to encode a
logical qubit that possesses superior coherence properties.  We have also
discussed how such logical qubits can be manipulated and measured.

{\it Note added:}~After finishing this work,
we became aware of Ref.~\onlinecite{GK} 
in which interbit couplings are also
proposed to reduce decoherence in a model Hamiltonian.

\begin{acknowledgments}
We thank J.S. Tsai, Y.X. Liu, and H. Fu for discussions. This work
was
supported in part by the NSA and ARDA under AFOSR contract
No.~F49620-02-1-0334, and by the NSF grant No.~EIA-0130383.  X.H. was
partially supported by ARDA and ARO at the University at Buffalo. J.Q.Y. was
also supported by the NSFC grant No.~10474013 and PCSIRT. 

\end{acknowledgments}



\begin{thebibliography}{99}

%

\bibitem{NEC1} Y. Nakamura, Yu. A. Pashkin, and J.S. Tsai, Nature (London) {\bf 398}, 
786 (1999).


\bibitem{MIT1} J.E. Mooij, T.P. Orlando, L. Levitov, L. Tian, C.H. van der Wal,
and S. Lloyd, Science {\bf 285}, 1036 (1999).


\bibitem{SUNY} J.R. Friedman, V. Patel, W. Chen, S.K. Tolpygo,
and J.E. Lukens, Nature (London) {\bf 406}, 43 (2000).


\bibitem{HAN} Y. Yu, S.Y. Han, X. Chu, S.I. Chu, and Z. Wang,
Science {\bf 296}, 889 (2002); J.M. Martinis,
S. Nam, J. Aumentado, and C. Urbina, Phys. Rev. Lett. {\bf 89}, 117901 (2002);
%
A.J. Berkley, H. Xu, R.C. Ramos, M.A. Gubrud, F.W. Strauch,
P.R. Johnson, J.R. Anderson, A.J. Dragt, C.J. Lobb,
and F.C. Wellstood, Science {\bf 300}, 1548 (2003).


\bibitem{MSS1} Yu. Makhlin, G. Sch{\" o}n, and A. Shnirman, 
Rev. Mod. Phys. {\bf 73}, 357 (2001).


\bibitem{SET} A.B. Zorin, F.J. Ahlers, J. Niemeyer, T. Weimann, H. Wolf 
V.A. Krupenin and S.V. Lotkhov, 
Phys. Rev. B {\bf 53}, 13682 (1996).


\bibitem{NEC2} Y. Nakamura, Yu.A. Pashkin, T. Yamamoto, and J.S. Tsai,
Phys. Rev. Lett. {\bf 88}, 047901 (2002).


\bibitem{PALA} E. Paladino, L. Faoro, G. Falci, and R. Fazio,
Phys. Rev. Lett. {\bf 88}, 228304 (2002).


\bibitem{SAC2} D. Vion, A. Aassime, A. Cottet, P. Joyez, H. Pothier, C. Urbina,
D. Esteve, and M.H. Devoret, Science {\bf 296}, 886 (2002).

\bibitem{COTT} A. Cottet, PhD thesis, University of Paris VI, 2002.

\bibitem{NEC3} Yu. A. Pashkin, T. Yamamoto, O. Astafiev, Y. Nakamura, D.V. Averin,
and J.S. Tsai, Nature (London) {\bf 421}, 823 (2003);
T. Yamamoto, Yu. A. Pashkin, O. Astafiev, Y. Nakamura, and J.S. Tsai,
{\it ibid.} {\bf 425}, 941 (2003).


\bibitem{YTN} J.Q. You, J.S. Tsai, and F. Nori, 
Phys. Rev. Lett. {\bf 89}, 197902 (2002); cond-mat/0306203.

\bibitem{ZR} 
G.M. Palma, K. A. Suominen, and A. K. Ekert, Proc. R. Soc. London, 
Ser. A {\bf 452}, 567 (1996);
P. Zanardi and M. Rasetti, 
Phys. Rev. Lett. {\bf 79}, 3306 (1997);
%
D.A. Lidar, I.L. Chuang, and K.B. Whaley, 
{\it ibid.} {\bf 81}, 2594 (1998).

\bibitem{DUAN}
L.M. Duan and G.C. Guo, 
Phys. Rev. A {\bf 57}, 737 (1998).


\bibitem{MB} F. Marquardt and C. Bruder, 
Phys. Rev. B {\bf 63}, 054514 (2001).

\bibitem{NOTE}
The fluxes through the dc-SQUID loops and the applied microwaves for quantum
operations will also produce decoherence.  However, the flux noise is much
weaker than the dominant background charge fluctuations in producing
decoherence.
%
The microwaves used for gate operations can be applied to the logical qubit
via the gate capacitances.  Thus their noise only produces gate voltage
fluctuations, whose effects are suppressed in the encoded logical qubit.



\bibitem{iontrap} J.I. Cirac and P. Zoller, Phys. Rev. Lett. {\bf 74}, 
4091 (1995).


\bibitem{GOV} M. Governale, M. Grifoni, and G. Sch\"{o}n,
Chem. Phys. {\bf 268}, 273 (2001).


\bibitem{STO} M.J. Storcz and F.K. Wilhelm,
Phys. Rev. A {\bf 67}, 042319 (2003).

\bibitem{THO} M. Thorwart and P. H\"{a}nggi, Phys. Rev. A {\bf 65},
012309 (2001).


\bibitem{Zhou} X.X. Zhou, M. Wulf, Z.W. Zhou, G.C. Guo, and M.J. Feldman, 
Phys. Rev. A {\bf 69}, 030301(R) (2004). 

\bibitem{NEC4} O. Astafiev, Yu.A. Pashkin, Y. Nakamura, T. Yamamoto, and J.S. Tsai,  
Phys. Rev. Lett. {\bf 93}, 267007 (2004).


\bibitem{GK} I.A. Grigorenko and D.V. Khveshchenko, Phys. Rev. Lett. {\bf 94},
040506 (2005).

\end{thebibliography}
\end{document}